\documentclass[english,preprint,aps]{revtex4}
\usepackage{graphicx}

\begin{document}
\title{Interface resistances and AC transport in a Luttinger liquid.}
\author{K.-V. Pham}
\address{Laboratoire de Physique des Solides, Universit\'{e} Paris-Sud, 91405 Orsay,\\
France.}
\date{\today}
\begin{abstract}
We consider a Luttinger liquid (LL) connected to two reservoirs  when the
two sample-reservoir interface resistances $R_{S}$ and $R_{D}$ are arbitrary
(not necessarily quantized at half-the-quantum of resistance). We compute
exactly the dynamical impedance of a Luttinger liquid and generalize earlier
expressions for its dynamical conductance in the following situations. (i)
We first consider a gated Luttinger liquid. It is shown that the Luttinger
liquid parameters $u$ and $K$ and the interface resistances $R_{S}$ and $%
R_{D}$ can be experimentally determined by measuring both the dynamical
conductance and impedance of a gated wire at second order in frequency. The
parallel law addition for the charge relaxation resistance $R_{q}$ is
explicitly recovered for these non-trivial interface resistances as $%
R_{q}^{-1}=R_{S}^{-1}+R_{D}^{-1}$. (ii) We discuss the AC response when only
one electrode is connected to the LL. (iii) Thirdly we consider application
of an arbitrary AC electric field along the sample and compute the dynamical
response of the LL with arbitrary interface resistances. The discussion is
then specialized to the case of a uniform electric field.
\end{abstract}
\maketitle

\section{Introduction.}

The Luttinger liquid (LL) is one of the best understood strongly correlated
system and departs strikingly from the more familiar Landau Fermi liquid
with features such as spin-charge separation and charge fractionalization
\cite{1}. Of interest is the exploration of Luttinger physics in a
mesoscopic context: several materials at the mesoscopic scale such as the
quantum wires or the carbon nanotubes have been indeed proposed as
realizations of a Luttinger liquid\cite{2}. In this regard AC transport
probes are an important tool because they allow access to the
non-equilibrium physics of the LL.

In this paper we propose to discuss AC transport in a Luttinger liquid
resistively connected to two reservoirs through arbitrary interface
resistances (not necessarily quantized at half a quantum of resistance): we
will consider in turn (i) a gated Luttinger liquid, (ii) the response when
one disconnects one of the reservoirs and (iii) application of an arbitrary
AC electric field along the sample.

The first calculation for settings (i) and (iii) was done by Ponomarenko
using the inhomogeneous Luttinger liquid model, where one models the
reservoirs as 1D non-interacting Fermi systems\cite{3}. His results for the
gated wire were later recovered using RPA by Blanter, Hekking and B\"{u}%
ttiker\cite{4} and using the inhomogeneous LL model and then a boundary
conditions formalism by Safi\cite{5}; the case of a constant electric field
was also considered by Sablikov and Shchamkhalova with results consistent
with Ponomarenko's; appealing to Shockley's theorem they however claim that
the real current measured at an electrode is not given by the electron
current but add a displacement current contribution due to a charging of the
reservoir caused by the charging of the wire itself: for a uniform electric
field this results in a net current measured equal to a spatial average of
the current through the wire\cite{6}. Another approach has been advocated by
Cuniberti, Sasseti and Kramer who consider an infinite system with
long-range interactions and compute an absorptive conductance which has the
advantage of being measurable by absorption of electromagnetic radiations
without application of voltage or current probes\cite{7}. An interesting
development in several of these groups has been a focus on both capacitive
and inductive effects with consideration of the kinetic inductance of a LL 
\cite{7,8}. A transmission line approach to AC transport in a LL has also
been proposed by Burke\cite{9} to investigate the plasmon physics of LL
based on earlier works on single-walled\cite{10} and multi-walled carbon
nanotubes\cite{11}; the LL is modelled as a RL line coupled capacitively to
a ground voltage. Additionally Burke discusses plasmon damping, a topic
rather unexplored so far in that context. Inclusion of Coulomb interactions
has also been considered in several papers\cite{7,8}.

In the case of short-range interactions (the pure LL) it is noteworthy that
the DC limit of earlier calculations corresponds to contact resistances
quantized at $R_{0}=e^{2}/h$ the quantum of resistance (or equivalently to
interface sample-electrode resistances $R_{0}/2$). We consider in this paper
a more general situation by allowing for interface resistances distinct from 
$R_{0}/2$: while leaving open the experimental possibility that interface
resistances are quantized at $R_{0}/2$, this permits dirty contacts to
electrodes, which a priori is a not too unreasonable assumption.

We generalize earlier expressions for the dynamical conductance in
situations (i) and (iii) above\cite{3,4,5}, and additionally compute the
dynamical impedance. Case (ii) where only one electrode is connected is
considered because it is a setting paradigmatic of time-dependent transport
where the role of displacement currents is especially clear. For case (i) we
show explicitly that to second order in frequency the LL can be represented
by an equivalent electrical circuit comprised of interface resistances $R_{S}
$ and $R_{D}$ connected in series to an intrinsic inductance (per unit
length) ${\cal L}=\frac{h}{2u\ Ke^{2}}$ (which is not purely kinetic but
includes the effect of interactions), the whole being capacitively coupled
to the ground through an intrinsic LL capacitance (per unit length) ${\cal C=%
}\frac{2Ke^{2}}{h\,u}$. (see Figure 1). This shows explicitly from first
principles the validity of the transmission line analogy considered by Burke
up to order two in a low-frequency expansion\cite{9} (our results however do
not assume Galilean invariance which implies in turn the relation $v_{F}=u\,K
$ for the Fermi velocity). 

This in turn shows that AC measurements of both the dynamical gate
conductance $G_{33}$ and the impedance of the system up to order two in
frequency allow full determination of the Luttinger liquid parameters $u$
and $K$ (the plasmon velocity and interaction strength) and of the interface
resistances $R_{S}$ and $R_{D}$. In particular the expected parallel law
addition for the charge relaxation resistance is explicitly recovered as $%
R_{q}^{-1}=R_{S}^{-1}+R_{D}^{-1}$.

The paper is organized as follows: in Section II, we introduce a boundary
condition formalism which allows for the modelling of reservoirs resistively
contacted to the Luttinger liquid. III. We discuss the gated Luttinger
liquid computing both dynamical conductance and impedance matrices, as well
as the LL connected to a single reservoir. IV. We impose an AC electric
field.

\section{Voltage drops at interfaces: modelling contact resistances through
boundary conditions.}

\subsection{\protect\bigskip Chiral chemical potential operators.}

We consider the standard Luttinger Hamiltonian for a wire of length $L=2a$. 
\[
H=\int_{-a}^{a}dx\ \frac{hu}{2K}\left( \rho _{+}^{2}+\rho _{-}^{2}\right)
+eV_{3}\left( \rho _{+}+\rho _{-}\right) 
\]

$V_{3}$ is a gate potential which controls the Fermi level of the LL, $\rho
_{+}$ and $\rho _{-}$ are chiral particle densities which obey the relation $%
\rho _{\pm }(x,t)=\rho _{\pm }(x\mp ut)$. Their sum is just the total
particle density $\rho -\rho _{0}$ while the electrical current is simply $%
i(x,t)=eu\left( \rho _{+}-\rho _{-}\right) $.

We now define the following operators:

\[
\mu _{\pm }(x,t)=\frac{\delta H}{\delta \rho _{\pm }(x,t)}
\]
Physically they correspond to (canonical) chemical potential operators:
their average value yields the energy needed to add a particle locally at
position $x$ to the chiral density: $\rho _{\pm }\longrightarrow \rho _{\pm
}+\delta (x)$. Similar operators have been introduced in Safi's boundary
conditions formalism\cite{5}: the main difference being that we consider
chiral chemical potentials linked to the eigenmodes of the Luttinger liquid
while she defines chemical potentials related to the left or right moving
(bare) electrons. Such chiral operators are much more convenient for
calculations since they are directly related to the LL eigenmodes.

From their definition it follows that:
\[
\mu _{\pm }(x,t)=\frac{hu}{K}\rho _{\pm }(x,t)+eV_{3}(t)
\]
and therefore:
\begin{equation}
i(x,t)=K\frac{e}{h}\left( \mu _{+}(x,t)-\mu _{-}(x,t)\right)
\label{current-def}
\end{equation}
It is convenient to redefine the chemical potentials by taking $V_{3}$ as
reference:
\begin{equation}
\mu _{\pm }^{\prime }(x,t)=\mu _{\pm }(x,t)-eV_{3}(t)  \label{shifted-chem}
\end{equation}
and using the fact that these shifted operators have a chiral time
evolution:
\[
\mu _{\pm }^{\prime }(x,t)=\mu _{\pm }^{\prime }(x\mp ut)
\]
it follows immediately that:

\begin{eqnarray}
\mu _{+}^{\prime }(a,\omega )& =\exp i\phi \ \mu _{+}^{\prime }(-a,\omega ),
\label{chemical} \\
\mu _{-}^{\prime }(a,\omega )& =\exp -i\phi \ \mu _{-}^{\prime }(-a,\omega ),
\nonumber \\
\phi =\omega \frac{2a}{u}.  \nonumber
\end{eqnarray}

\subsection{\protect\bigskip Interface resistances as boundary conditions.}

Up to now the Luttinger liquid is free standing. In real experimental
settings coupling to probes is however unavoidable but in the absence of an
exact solution of the problem of a mesoscopic LL wire coupled to many
electrodes we decide to model the coupling to reservoirs through boundary
conditions imposed on the otherwise free standing Luttinger liquid.

To enforce that we appeal to Sharvin-Imry contact resistance\cite{12}: at
the interface between a reservoir and a ballistic wire there is a voltage
drop between the electrode voltage and the mean chemical potential within
the wire; for a two-terminal geometry this in turn implies the existence of
a contact resistance which can be viewed as the series addition of two
interface resistances. For constrictions adiabatically connected to the
reservoirs the contact resistance is quantized as $R_{0}=h/e^{2}$. But in
general it need not be; as shown by B\"{u}ttiker incoherent transport
through barriers can affect quantization\cite{13}.

For the LL we therefore make the hypothesis that as far as transport is
concerned the resistive coupling to reservoirs can be modelled as:

\begin{eqnarray*}
i(-a,t) &=&\frac{1}{R_{S}}\left( V_{S}(t)-\frac{\mu _{+}(-a,t)+\mu _{-}(-a,t)%
}{2e}\right) \\
i(a,t) &=&\frac{1}{R_{D}}\left( \frac{\mu _{+}(a,t)+\mu _{-}(a,t)}{2e}%
-V_{D}(t)\right)
\end{eqnarray*}

In the above equations we have considered two electrodes connected at the
boundaries of the LL, the left electrode being a source at voltage $V_{S}(t)$
and the right electrode being a drain at voltage $V_{D}(t)$. Currents are
oriented from left to right.

We stress that these relations are operator ones: we work therefore in the
Heisenberg representation. For computation of noise properties it is indeed
crucial that these relations are enforced at the operator level and not as
average; knowledge of a current average is insufficient to specify
fluctuation properties.

These relations extend an earlier formalism developed by the author and
collaborators\cite{14}: the main difference is that earlier we considered
the chemical potentials as uniform (as is the case in a DC context) in a
grand-canonical approach while here we work in a canonical setting with
local potentials, which is more suitable to the AC context.

We note in passing that such relations can be derived explicitly in several
exactly solvable models: for instance for the inhomogeneous Luttinger liquid
with interface resistances $R_{S}=R_{D}=R_{0}/2$; for a chiral Luttinger
liquid connected by a point contact to a Fermi liquid with $R=R_{0}/2$ or
more generally for a reservoir which is a LL with LL parameter $K_{res}$
also connected through a point contact to the sample, the interface
resistance is $R=R_{0}/2K_{res}$. Safi's boundary conditions\cite{5} $%
V_{S/D}(t)=\frac{\delta H}{\delta \rho _{\pm }^{0}(x,t)}$ where $\rho _{\pm
}^{0}$ are the {\it electron densities at the right and left Fermi points
(and not the chiral densities)} coincide with our boundary conditions for
the special values $R_{S}=R_{D}=R_{0}/2$. For a detailed discussion of these
relations we refer to our earlier work \cite{14}.

Simple though these relations may seem they permit to go beyond earlier  AC
results found by using for instance the inhomogeneous LL model as will next
be shown.

\section{\protect\bigskip Dynamical response of a gated wire.}

\subsection{\protect\bigskip Dynamical impedance.}

We now consider time-varying voltage sources $V_{S}(t)=V_{1}\exp i\omega t$
and $V_{D}(t)=V_{2}\exp i\omega t$ and a gate voltage $V_{3}(t)=V_{3}\exp
i\omega t$ and compute the dynamical impedance and conductance matrices of
the LL. The currents at the boundaries of the system then acquire the same
time dependence; we define currents as entering the system: 
\[
\left( 
\begin{array}{c}
i_{1} \\ 
i_{2}
\end{array}
\right) =\left( 
\begin{array}{c}
i(-a,\omega ) \\ 
-i(a,\omega )
\end{array}
\right) . 
\]

To enforce current conservation there will in general be a displacement
current corresponding to the charging of the sample. In that section we fix
the currents at the boundaries as $i_{1}=i_{1}^{0}\exp i\omega t$ and $%
i_{2}=i_{2}^{0}\exp i\omega t$. Therefore the voltages $V_{S}(t)=V_{1}\exp
i\omega t$ and $V_{D}(t)=V_{2}\exp i\omega t$ can be viewed as responses to
the currents.

The boundary conditions are therefore rewritten as: 
\begin{equation}
\left( 
\begin{array}{c}
i_{1} \\ 
i_{2}
\end{array}
\right) =\left( 
\begin{array}{c}
\frac{1}{R_{S}}\left( V_{1}-\frac{\mu _{+}(-a)+\mu _{-}(-a)}{2e}\right) \\ 
\frac{1}{R_{D}}\left( V_{2}-\frac{\mu _{+}(-a)+\mu _{-}(-a)}{2e}\right)
\end{array}
\right)  \label{boundary}
\end{equation}
Using eq.(\ref{current-def}) and eq.(\ref{chemical}) it follows that: 
\begin{equation}
\left( 
\begin{array}{c}
i_{1} \\ 
i_{2}
\end{array}
\right) =\frac{Ke}{h}\left( 
\begin{array}{cc}
1 & -\exp i\phi \\ 
-\exp i\phi & 1
\end{array}
\right) \left( 
\begin{array}{c}
\mu _{+}^{\prime }(-a) \\ 
\mu _{-}^{\prime }(a)
\end{array}
\right)  \label{current-mu}
\end{equation}
Defining the vector $\overrightarrow{\mu }$ as: 
\begin{equation}
\overrightarrow{\mu }=\left(
\begin{array}{c}
\mu _{+}^{\prime }(-a) \\ 
\mu _{-}^{\prime }(a)
\end{array}
\right) ,  \label{mu}
\end{equation}
and using eq.(\ref{chemical},\ref{boundary},\ref{shifted-chem}) there
follows: 
\begin{eqnarray*}
\left( 
\begin{array}{c}
V_{1}-V_{3} \\ 
V_{2}-V_{3}
\end{array}
\right) &=&\left( 
\begin{array}{c}
\frac{\mu _{+}^{\prime }(-a)+\mu _{-}^{\prime }(-a)}{2e} \\ 
\frac{\mu _{+}^{\prime }(-a)+\mu _{-}^{\prime }(-a)}{2e}
\end{array}
\right) +\left( 
\begin{array}{cc}
R_{S} & 0 \\
0 & R_{D}
\end{array}
\right) \left( 
\begin{array}{c}
i_{1} \\ 
i_{2}
\end{array}
\right) \\
&=&\frac{1}{e}\left( 
\begin{array}{cc}
\frac{1}{2} & \frac{1}{2}\exp i\phi \\ 
\frac{1}{2}\exp i\phi & \frac{1}{2}
\end{array}
\right) \overrightarrow{\mu }+\left( 
\begin{array}{cc}
R_{S} & 0 \\ 
0 & R_{D}
\end{array}
\right) \left( 
\begin{array}{c}
i_{1} \\ 
i_{2}
\end{array}
\right) .
\end{eqnarray*}
Inserting eq.(\ref{current-mu}), there comes: 
\begin{equation}
\left( 
\begin{array}{c}
V_{1}-V_{3} \\ 
V_{2}-V_{3}
\end{array}
\right) =\frac{1}{e}\left( 
\begin{array}{cc}
\frac{1}{2}+K\overline{R_{S}} & \exp i\phi \left( \frac{1}{2}-K\overline{%
R_{S}}\right) \\ 
\exp i\phi \left( \frac{1}{2}-K\overline{R_{D}}\right) & \frac{1}{2}+K%
\overline{R_{D}}
\end{array}
\right) \overrightarrow{\mu }  \label{v-mu}
\end{equation}
where $\overline{R_{D}}=R_{D}/R_{0}$ and $\overline{R_{S}}=R_{S}/R_{0}$ are
resistances measured against the quantum of resistance $R_{0}=\frac{h}{e^{2}}
$. Inverting now eq.(\ref{current-mu}) and inserting it in eq.(\ref{v-mu})
one gets: 
\[
\left( 
\begin{array}{c}
V_{1}-V_{3} \\ 
V_{2}-V_{3}
\end{array}
\right) =\frac{h}{Ke^{2}}\left( 
\begin{array}{cc}
\frac{1}{2}+K\overline{R_{S}} & \exp i\phi \left( \frac{1}{2}-K\overline{%
R_{S}}\right) \\ 
\exp i\phi \left( \frac{1}{2}-K\overline{R_{D}}\right) & \frac{1}{2}+K%
\overline{R_{D}}
\end{array}
\right) \left( 
\begin{array}{cc}
1 & -\exp i\phi \\ 
-\exp i\phi & 1
\end{array}
\right) ^{-1}\left(
\begin{array}{c}
i_{1} \\ 
i_{2}
\end{array}
\right) 
\]

Defining the dynamical impedance matrix as: 
\[
\left( 
\begin{array}{c}
V_{1}-V_{3} \\ 
V_{2}-V_{3}
\end{array}
\right) =\underline{{\Huge Z}}\left( 
\begin{array}{c}
i_{1} \\
i_{2}
\end{array}
\right) ,
\]
one finally finds: 
\begin{equation}
\underline{{\Huge Z}}=\left( 
\begin{array}{cc}
R_{S}+i\frac{R_{0}}{2K}\cot \phi  & i\frac{R_{0}}{2K\sin \phi } \\ 
i\frac{R_{0}}{2K\sin \phi } & R_{D}+i\frac{R_{0}}{2K}\cot \phi 
\end{array}
\right) .  \label{impedance-matrix}
\end{equation}
where $\phi =\omega \frac{L}{u}$ ($L$ is the length of the system, and $u$
is the plasmon velocity). This is the main result of this sub-section.

\subsection{\protect\bigskip Intrinsic inductance of the Luttinger liquid.}

We now consider the following experimental arrangement in order to measure
the impedance of the LL: 
\[
i_{1}=-i_{2}=i_{0}\exp i\omega t 
\]
.

The impedance of the system is therefore related to the matrix elements of
the full impedance matrix by: 
\[
Z=\frac{V_{1}-V_{2}}{i_{1}}=\underline{{\Huge Z}}_{11}+\underline{{\Huge Z}}%
_{22}-\underline{{\Huge Z}}_{12}-\underline{{\Huge Z}}_{21} 
\]
and therefore: 
\[
Z=R_{S}+R_{D}-i\frac{R_{0}}{K}\tan \left( \frac{\phi }{2}\right) . 
\]

That especially simple formula admits as low frequency limit: 
\[
Z=R_{S}+R_{D}-i\frac{R_{0}}{K}\frac{\omega L}{2u}+i\frac{R_{0}}{3K}\left( 
\frac{\omega L}{2u}\right) ^{3}+{\cal O}(\omega ^{3}) 
\]
where $L=2a$ is the size of the system.

Comments:

(i) This shows firstly that the total contact resistance results as it
should be from a series addition of the two interface resistances $R_{S}$
and $R_{D}$.

(ii) Secondly, since $Z=R_{S}+R_{D}-i\omega \left( {\cal L}L\right) +{\cal O}%
(\omega )$ there appears an inductance per unit length:

\bigskip
\[
{\cal L}=\frac{h}{2u\ Ke^{2}}. 
\]

This is as it should be; indeed direct inspection of the Luttinger
Hamiltonian shows that the Luttinger liquid must have an inductance
precisely set at that value. Indeed: 
\[
H=\int_{-a}^{a}dx\ \frac{hu}{4K}\rho _{{}}^{2}+\frac{hu\,K}{4}\overline{j}%
_{{}}^{2}
\]
where $\overline{j}=\rho _{+}^{0}-\rho _{-}^{0}$ is the difference between
bare right and left electron densities (at right and left Fermi points $\pm
k_{F}$). Rewriting the Hamiltonian in terms of charge density and current: 
\[
\rho _{e}=e\ \rho ;\;j_{e}=e\ u\,K\overline{\,j}
\]
(the last expression follows from charge conservation and the equations of
motion) there follows: 
\[
H=\int_{-a}^{a}dx\ \frac{hu}{4Ke^{2}}\rho _{e}^{2}+\frac{h}{4u\,Ke^{2}}%
j_{e}^{2}\text{.}
\]
This shows indeed an inductance per unit length ${\cal L}=\frac{h}{2u\ Ke^{2}%
}$ while the zero mode of the first term yields $\frac{hu}{4Ke^{2}L}%
Q_{{}}^{2}$ which shows there is a capacitance per unit length: 
\[
{\cal C=}\frac{2Ke^{2}}{h\,u}.
\]
While there has been ample emphasis on the intrinsic capacitance of the
Luttinger liquid\cite{15,4} the fact that the LL possesses an intrinsic
inductance is less well stressed: see however\cite{7,10}. We note in passing
that the term $\frac{h}{4u\,Ke^{2}}j_{e}^{2}$ in the Hamiltonian results
from both kinetic energy and interactions: the inductance ${\cal L}=\frac{h}{%
2u\ Ke^{2}}$ has therefore a mixed origin and is not merely contrarily to what Burke argues a kinetic
inductance \cite{9}: this point is somewhat obscured by
the fact that in Galilean invariant systems $v_{F}=u\,K$ which implies then
that the intrinsic LL inductance assumes exactly the same value as in a
non-interacting system; however since Galilean invariance is in general not
realized the previous identity does not hold and a renormalization by the
interactions of the kinetic inductance should follow. At any rate whether
experiments can provide independent measurements
 of both $u$ and $K$: therefore one need not assume that  $v_{F}=u\,K$, since the
 validity or non-validity of that relation can be checked.

(iii) For carbon nanotubes assuming a length $L\sim 1\mu m$ and a plasmon
velocity of the order of $v_{F}=10^{5}ms^{-1}$ means that each successive
term in the low-frequency expansion of the impedance goes as $R_{0}\left( 
\frac{\omega }{100GHz}\right) ^{n}$. This implies that at already a
frequency of about $10kHz$ the inductive correction is $\delta Z/Z=10^{-7}$.
While the first order correction is quite measurable the next order (three)
correction is much less accessible unless one goes to the $GHz$ range.

\subsection{Dynamical conductance.}

We now fix voltages; therefore the relation $i_{1}=-i_{2}$ is not valid any
more. As amply stressed by B\"{u}ttiker there is a displacement current $%
i_{3}$ due to the charging of the system\cite{16}. Current conservation is
enforced only if one considers that additional current.

One now inverts the relation $\left( 
\begin{array}{c}
V_{1}-V_{3} \\ 
V_{2}-V_{3}
\end{array}
\right) =\underline{Z}\left( 
\begin{array}{c}
i_{1} \\ 
i_{2}
\end{array}
\right) $; whence the upper $2\times 2$ part of the dynamical conductance
matrix: 
\[
\left( 
\begin{array}{c}
i_{1} \\ 
i_{2}
\end{array}
\right) =\frac{KG_{0}}{\left( \frac{1}{2}+K\overline{R_{S}}\right) \left( 
\frac{1}{2}+K\overline{R_{D}}\right) -\exp i2\phi \left( \frac{1}{2}-K%
\overline{R_{S}}\right) \left( \frac{1}{2}-K\overline{R_{D}}\right) }\times 
\]
\[
\times \left( 
\begin{array}{cc}
\frac{1}{2}+K\overline{R_{D}}+\exp i2\phi \left( \frac{1}{2}-K\overline{R_{D}%
}\right) & -\exp i\phi \\ 
-\exp i\phi & \frac{1}{2}+K\overline{R_{S}}+\exp i2\phi \left( \frac{1}{2}-K%
\overline{R_{S}}\right)
\end{array}
\right) \left( 
\begin{array}{c}
V_{1}-V_{3} \\ 
V_{2}-V_{3}
\end{array}
\right) 
\]

Using current conservation $\sum_{k}i_{k}=0$ the full conductance matrix
follows: 
\[
\left(
\begin{array}{c}
i_{1} \\ 

i_{2} \\ 
i_{3}
\end{array}
\right) =\underline{{\Huge G}}\left( 
\begin{array}{c}
V_{1} \\ 
V_{2} \\ 
V_{3}
\end{array}
\right) 
\]
with matrix elements: 
\[
\underline{{\Huge G}}_{11}=\frac{KG_{0}\left[ \frac{1}{2}+K\overline{R_{D}}%
+\exp i2\phi \left( \frac{1}{2}-K\overline{R_{D}}\right) \right] }{\left( 
\frac{1}{2}+K\overline{R_{S}}\right) \left( \frac{1}{2}+K\overline{R_{D}}%
\right) -\exp i2\phi \left( \frac{1}{2}-K\overline{R_{S}}\right) \left( 
\frac{1}{2}-K\overline{R_{D}}\right) } 
\]
\[
\underline{{\Huge G}}_{12}=\underline{{\Huge G}}_{21}=\frac{-KG_{0}\exp
i\phi }{\left( \frac{1}{2}+K\overline{R_{S}}\right) \left( \frac{1}{2}+K%
\overline{R_{D}}\right) -\exp i2\phi \left( \frac{1}{2}-K\overline{R_{S}}%
\right) \left( \frac{1}{2}-K\overline{R_{D}}\right) } 
\]
\[
\underline{{\Huge G}}_{22}=\frac{KG_{0}\left[ \frac{1}{2}+K\overline{R_{S}}%
+\exp i2\phi \left( \frac{1}{2}-K\overline{R_{S}}\right) \right] }{\left( 
\frac{1}{2}+K\overline{R_{S}}\right) \left( \frac{1}{2}+K\overline{R_{D}}%
\right) -\exp i2\phi \left( \frac{1}{2}-K\overline{R_{S}}\right) \left( 
\frac{1}{2}-K\overline{R_{D}}\right) } 
\]

\[
\underline{{\Huge G}}_{13}=\underline{{\Huge G}}_{31}=-\underline{{\Huge G}}%
_{11}-\underline{{\Huge G}}_{12} 
\]
\[
\underline{{\Huge G}}_{23}=\underline{{\Huge G}}_{32}=-\underline{{\Huge G}}%
_{22}-\underline{{\Huge G}}_{21} 
\]
\begin{eqnarray*}
\underline{{\Huge G}}_{33} &=&\underline{{\Huge G}}_{11}+\underline{{\Huge G}%
}_{22}-\underline{{\Huge G}}_{12}-\underline{{\Huge G}}_{21} \\
&=&\frac{KG_{0}\left[ 1+K\left( \overline{R_{S}}+\overline{R_{D}}\right)
-2\exp i\phi +\exp i2\phi \left( 1-K\left( \overline{R_{S}}+\overline{R_{D}}%
\right) \right) \right] }{\left( \frac{1}{2}+K\overline{R_{S}}\right) \left( 
\frac{1}{2}+K\overline{R_{D}}\right) -\exp i2\phi \left( \frac{1}{2}-K%
\overline{R_{S}}\right) \left( \frac{1}{2}-K\overline{R_{D}}\right) }
\end{eqnarray*}
\newline

A useful check is to set the interface resistances to $R_{S}=R_{D}=R_{0}/2$:
one recovers immediately eq.(10-11) of Blanter et al\cite{4}.

We now expand the gate conductance $\underline{G}_{33}$: 
\begin{equation}
\underline{{\Huge G}}_{33}=-i{\cal C}L\omega +\omega ^{2}\left( {\cal C}%
L\right) ^{2}R_{q}+i\omega ^{3}\left( {\cal C}L\right) ^{3}R_{q}^{2}\left( 1+%
\frac{R_{0}^{2}}{4K^{2}R_{q}R_{C}}-\frac{R_{0}^{2}}{12K^{2}R_{q}^{2}}\right)
+{\cal O}(\omega ^{3})  \label{gate}
\end{equation}
where ${\cal C=}\frac{2Ke^{2}}{h\,u}$, $R_{q}=(R_{S}R_{D})/(R_{S}+R_{D})$
and $R_{C}=R_{S}+R_{D}$ is the contact resistance.

\subsection{\protect\bigskip Discussion.}

The previous expression shows:

(i) firstly that the capacitance per unit length ${\cal C=}\frac{2Ke^{2}}{%
h\,u}$ is independent of the coupling to the reservoirs: this is quite
sensible; its value is just that expected from a direct inspection of the
Luttinger Hamiltonian (see above). Measurement of both ${\cal L}=\frac{h}{%
2u\ Ke^{2}}$ and ${\cal C=}\frac{2Ke^{2}}{h\,u}$ therefore provides a direct
way to get the values of $u$ and $K$. As already noticed by Burke using the
telegraphist equation\cite{9} the plasmon velocity is just $\sqrt{{\cal LC}}%
{\cal =}\frac{1}{u}$ an identity well-known to electrical engineers while
the transmission line impedance is just: $Z_{0}=\sqrt{{\cal L}/{\cal C}}%
{\cal =}\frac{1}{2K}$.

This justifies a posteriori the transmission line analogy proposed by Burke.
Note however that our results show that the transmission line analogy is
valid only up to order two in a low frequency expansion.

(ii) That capacitance ${\cal C}$ is fully chemical and corresponds to (the
inverse of) the density of states: as stressed by B\"{u}ttiker\cite{16} in a
general experimental setting one has to add a geometrical capacitance
describing the coupling to a wire so that the total capacitance is $%
C_{tot}^{-1}=(L{\cal C)}^{-1}+C_{geom}^{-1}$.

(iii) There appears a charge relaxation resistance $R_{q}$ which obeys a
parallel addition law: $R_{q}^{-1}=R_{S}^{-1}+R_{D}^{-1}$. This is quite
sensible because relaxation probabilities should add for independent
relaxation processes. The charge relaxation resistance is distinct from the
contact resistance in that it corresponds to an $RC$ time for the
discharging of a system and not (directly) to energy dissipation\cite{16}.
We also observe that by measuring both impedance and gate conductance up to
order two in frequency one can directly measure both interface resistances:
there is therefore no need to assume that they are a priori set at $%
R_{S}=R_{D}=R_{0}/2$ since this can be checked experimentally.

(iv) Several approaches have been advocated for determining the Luttinger
parameters using AC measurements. Ponomarenko\cite{3}, Sablikov and
Shchamkhalova\cite{6} proposed to measure the period of the conductance
since they are oscillating functions of the parameter $\phi =\omega \frac{L}{%
u}$. This has been criticized by Blanter et al. who argue that the frequency
is quite high (GHz range)\cite{4}. Safi proposed to measure the conductance
at low frequency and measure its deviation to the DC limit\cite{5}, by
showing that at low frequency for a symmetric electric field arrangement one
can neglect the displacement current so that $G_{12}=-G_{11}=G_{0}(1+i\omega 
\frac{L}{2u\,K})$: this allows access to the product $u\,K$.\thinspace\
Blanter et al. argued that such deviations are hard to identify and proposed
to measure the gate conductance up to order three\cite{4} to determine the
values of $u$ and $K$.

As is apparent from our discussion of the inductance of the Luttinger liquid
a joint measurement of both impedance and gate conductance circumvents the
need to go to order three in frequency: it is sufficient to go to order one,
which means that measurements at the kHz range should be enough rather than
the 100 GHz range. That measurement of the impedance is equivalent to the
measurement of $G_{12}$ proposed by Safi with the advantage that fixing the
currents as in an impedance measurement avoids the complications of
displacement current. In addition note that our derivation does not require
a symmetric configuration for the electrodes.

(v) But if conversely one is able to make measurements up to order three
(i.e. up to $G_{0}\left( \frac{\omega }{100GHz}\right) ^{3}$ , or at $GHz$
range), the order third term instead of providing a {\it mere fit} of the LL
theory to experiments now constitutes a distinct non-trivial prediction of
the theory. But as is obvious from eq.(\ref{gate}) instead of the simpler
expression $i\omega ^{3}\left( {\cal C}L\right) ^{3}R_{q}^{2}\left( 1-\frac{1%
}{3K^{2}}\right) $ given by Blanter et al., charge relaxation and contact
resistances enter in an intricate way if they do not take the trivial values 
$Rq=h/4e^{2}$ and $R_{C}=h/e^{2}$. The same comment applies to the dynamical
impedance: 
\[
Z=R_{S}+R_{D}-i\omega L{\cal L}+i\omega ^{3}L^{3}\frac{{\cal L}^{2}{\cal C}}{%
12}+{\cal O}(\omega ^{3}).
\]
We note in passing that the appearance of $K^{2}$ in the formula can be
understood quite simply as the translation of inductive corrections since $%
{\cal L}/{\cal C=}\frac{R_{0}^{2}}{4K^{2}}$ .

\subsection{AC response of a LL connected to a single reservoir.}

That experimental setting is especially interesting because in the DC limit
there is no current. The dynamics within an AC experiment is wholly governed
by the charge dynamics within the sample and illustrates nicely the role of
the displacement current as stressed by B\"{u}ttiker\cite{16}.

To the author's knowledge such a setting for a LL has not been treated in
the literature even in the case of $R_{S}=R_{0}/2$. Yet in our formalism
that situation is quite straightforwardly described: it suffices to take the
limit of infinite interface resistance $R_{D}$ if one wants for instance to
disconnect the drain electrode. No current can flow into the drain and the
current $i_{2}$ is therefore zero. The only non-zero matrix elements of the
dynamical conductance are $\underline{G}_{11},\underline{G}_{13},\underline{G%
}_{33},\underline{G}_{31}$ and are determined by a single number: 
\begin{eqnarray*}
\underline{{\Huge G}}_{11} &=&\underline{{\Huge G}}_{33}=\frac{KG_{0}\left(
1-\exp i2\phi \right) }{\left( \frac{1}{2}+K\overline{R_{S}}\right) +\exp
i2\phi \left( \frac{1}{2}-K\overline{R_{S}}\right) }, \\
\underline{{\Huge G}}_{13} &=&\underline{{\Huge G}}_{31}=-\underline{{\Huge G%
}}_{11}.
\end{eqnarray*}
In that situation the incoming current charges the Luttinger liquid and
therefore $i_{1}+i_{3}=0$ the displacement current compensates exactly the
charge current. The impedance $Z=\frac{V_{1}-V_{3}}{i_{1}}$ is the inverse
of $\underline{G}_{11}$: 
\begin{eqnarray*}
Z &=&R_{S}+\frac{i}{2K}\cot \phi  \\
&=&\frac{1}{-i{\cal C}L\omega }+R_{S}+{\cal O}(1).
\end{eqnarray*}

The low-frequency expansion of the conductance is:
\[
\underline{{\Huge G}}_{33}=-i{\cal C}L\omega +\omega ^{2}\left( {\cal C}%
L\right) ^{2}R_{S}+i\omega ^{3}\left( {\cal C}L\right) ^{3}R_{S}^{2}\left( 1-%
\frac{R_{0}^{2}}{12K^{2}R_{S}^{2}}\right) .
\]
In such a setting the inductive effects are much more difficult to see: one
must go to order three in the conductance to observe them.

\section{Dynamical response of a LL to an AC electric field.}

\subsection{\protect\bigskip Equations of motion and boundary conditions.}

We now apply an AC electric field along the sample and give therefore a
spatial dependence to $V_{3}$: 
\[
V_{3}(x,t)=V_{3}(x)\,e^{i\omega t}.
\]
The Luttinger Hamiltonian can be rewritten in terms of the phase field $%
\theta $ conjugate to the density as (we set $\hbar =1$): 
\[
H=\int_{-a}^{a}dx\ \frac{\pi u}{2K}\rho ^{2}+\frac{u\,K}{2\pi }\left(
\partial _{x}\theta \right) ^{2}+eV_{3}\rho 
\]
The equations of motion in the Heisenberg representation for the density and
the particle current are: 
\begin{eqnarray*}
\partial _{t}^{2}\rho -u^{2}\partial _{x}^{2}\rho  &=&\frac{\,u\,Ke}{\,\pi }%
\partial _{x}^{2}V_{3} \\
\partial _{t}^{2}j-u^{2}\partial _{x}^{2}j &=&-\frac{\,u\,Ke}{\,\pi }%
\partial _{x,t}^{2}V_{3}
\end{eqnarray*}
where the particle current is (as can be checked from the current
conservation equation): 
\[
j=\frac{-u\,K}{\pi }\partial _{x}\theta 
\]
These operators can therefore be written as: 
\begin{eqnarray*}
\rho (x,t) &=&\rho _{+}(t-\frac{x}{u})+\rho _{-}(t+\frac{x}{u})+\rho
_{0}(x)e^{i\omega t} \\
j(x,t) &=&j_{+}(t-\frac{x}{u})+j_{-}(t+\frac{x}{u})+j_{0}(x)\,e^{i\omega t}
\end{eqnarray*}
where $\rho _{0}(x)e^{i\omega t}$ and $j_{0}(x)\,e^{i\omega t}$ are
arbitrary particular solutions of the equations of motion. One can choose $%
\rho _{0}(x)$ to be proportional to $j_{0}(x)$. Indeed using current
conservation $\partial _{t}\rho +\partial _{x}j=0$ it follows immediately
that: 
\begin{eqnarray*}
\partial _{x}j &=&-\partial _{t}\rho =-\partial _{t}\left( \rho _{+}+\rho
_{-}\right) -i\omega \rho _{0}(x)e^{i\omega t} \\
&=&u\partial _{x}\left( \rho _{+}-\rho _{-}\right) -i\omega \rho
_{0}(x)e^{i\omega t}
\end{eqnarray*}
which implies that we can set: 
\[
j_{\pm }=\pm u\rho _{\pm },\qquad \rho _{0}=\frac{-1}{i\omega }\partial
_{x}j_{0.}
\]

The chiral chemical potential operators are now:\newline
\[
\mu _{\pm }(x,t)=\frac{2\pi u}{K}\rho _{\pm }(x,t)+eV_{3}(x,t)+\frac{\pi u}{K%
}\rho _{0}\pm \frac{\pi }{K}j_{0}. 
\]
It is readily checked that the current operator is given by eq.(\ref
{current-def}): 
\begin{equation}
i(x,t)=ej(x,t)=K\frac{e}{h}\left( \mu _{+}(x,t)-\mu _{-}(x,t)\right) .
\end{equation}
Again it is convenient to shift the chemical potential operators to have
operators which have a purely chiral time-evolution: 
\[
\mu _{\pm }^{\prime }(x,t)=\mu _{\pm }(x,t)-eV_{3}(t)-\frac{\pi u}{K}\rho
_{0}\mp \frac{\pi }{K}j_{0}. 
\]
We now consider the following boundary conditions: 
\begin{eqnarray*}
ej(-a,t) &=&\frac{1}{R_{S}}\left( V_{3}(-a,t)-\frac{\mu _{+}(-a,t)+\mu
_{-}(-a,t)}{2e}\right) , \\
ej(a,t) &=&\frac{1}{R_{D}}\left( \frac{\mu _{+}(a,t)+\mu _{-}(a,t)}{2e}%
-V_{3}(a,t)\right) .
\end{eqnarray*}
These boundary conditions correspond to source and drain voltages set to the
ground (zero voltage): therefore only the potential $V_{3}$ appears; it
corresponds to the energy gained due to the initial (or final) acceleration
given by the applied electric field. In the previous section one did not
have to take it into account since no electric field was applied.

\subsection{\protect\bigskip Dynamical response.}

Defining again the currents as entering the system:

\[
\left( 
\begin{array}{c}
i_{1} \\ 
i_{2}
\end{array}
\right) =\left( 
\begin{array}{c}
ej(-a,t) \\ 
-ej(a,t)
\end{array}
\right) , 
\]
the boundary conditions are rewritten as:

\[
\left( 
\begin{array}{c}
i_{1} \\ 
i_{2}
\end{array}
\right) =\left( 
\begin{array}{c}
\frac{1}{R_{S}}\left( -\frac{\pi u}{eK}\rho _{0}(-a)-\frac{\mu _{+}^{\prime
}(-a)+\mu _{-}^{\prime }(-a)}{2e}\right) \\ 
\frac{1}{R_{D}}\left( \frac{\pi u}{eK}\rho _{0}(a)-\frac{\mu _{+}^{\prime
}(-a)+\mu _{-}^{\prime }(-a)}{2e}\right)
\end{array}
\right) . 
\]
Therefore: 
\begin{eqnarray}
\left( 
\begin{array}{c}
-\frac{\pi u}{eK}\rho _{0}(-a) \\ 
\frac{\pi u}{eK}\rho _{0}(a)
\end{array}
\right) &=&\left( 
\begin{array}{c}
\frac{\mu _{+}^{\prime }(-a)+\mu _{-}^{\prime }(-a)}{2e} \\ 
\frac{\mu _{+}^{\prime }(-a)+\mu _{-}^{\prime }(-a)}{2e}
\end{array}
\right) +\left( 
\begin{array}{cc}
R_{S} & 0 \\ 
0 & R_{D}
\end{array}
\right) \left( 
\begin{array}{c}
i_{1} \\ 
i_{2}
\end{array}
\right)  \label{rho-mu} \\
&=&\frac{1}{e}\left( 
\begin{array}{cc}
\frac{1}{2} & \frac{1}{2}\exp i\phi \\ 
\frac{1}{2}\exp i\phi & \frac{1}{2}
\end{array}
\right) \overrightarrow{\mu }+\left( 
\begin{array}{cc}
R_{S} & 0 \\ 
0 & R_{D}
\end{array}
\right) \left( 
\begin{array}{c}
i_{1} \\ 
i_{2}
\end{array}
\right)  \nonumber
\end{eqnarray}

where the vector $\overrightarrow{\mu }$ is defined as above (eq.(\ref{mu}%
)). But according to eq.(\ref{current-def}) the current is rewritten as: 
\[
\left( 
\begin{array}{c}
i_{1} \\ 
i_{2}
\end{array}
\right) =\frac{Ke^{{}}}{h}\left( 
\begin{array}{cc}
1 & -\exp i\phi \\ 
-\exp i\phi & 1
\end{array}
\right) \overrightarrow{\mu }+\left( 
\begin{array}{c}
ej_{0}(-a) \\ 
-ej_{0}(a)
\end{array}
\right) . 
\]
Therefore substitution of the previous equation in eq.(\ref{rho-mu}) yields: 
\begin{equation}
\left( 
\begin{array}{c}
-\frac{\pi u}{eK}\rho _{0}(-a) \\ 
\frac{\pi u}{eK}\rho _{0}(a)
\end{array}
\right) =\frac{1}{e}\left( 
\begin{array}{cc}
\frac{1}{2}+K\overline{R_{S}} & \exp i\phi \left( \frac{1}{2}-K\overline{%
R_{S}}\right) \\ 
\exp i\phi \left( \frac{1}{2}-K\overline{R_{D}}\right) & \frac{1}{2}+K%
\overline{R_{D}}
\end{array}
\right) \overrightarrow{\mu }+\left( 
\begin{array}{c}
ej_{0}(-a)R_{S} \\ 
-ej_{0}(a)R_{D}
\end{array}
\right)  \label{mu-j}
\end{equation}
Elimination of $\overrightarrow{\mu }$ then yields: 
\begin{equation}
\left( 
\begin{array}{c}
i_{1} \\ 
i_{2}
\end{array}
\right) =\underline{{\Huge Z}}^{-1}\left( 
\begin{array}{c}
-\frac{\pi u}{eK}\rho _{0}(-a)-ej_{0}(-a)R_{S} \\ 
\frac{\pi u}{eK}\rho _{0}(a)+ej_{0}(a)R_{D}
\end{array}
\right) +\left( 
\begin{array}{c}
ej_{0}(-a) \\ 
-ej_{0}(a)
\end{array}
\right)  \label{main-result3}
\end{equation}

where the matrix $\underline{Z}$ is the same dynamical impedance matrix
found above in eq.(\ref{impedance-matrix}): 
\begin{equation}
\underline{{\Huge Z}}=\left( 
\begin{array}{cc}
R_{S}+i\frac{R_{0}}{2K}\cot \phi & i\frac{R_{0}}{2K\sin \phi } \\ 
i\frac{R_{0}}{2K\sin \phi } & R_{D}+i\frac{R_{0}}{2K}\cot \phi
\end{array}
\right) .
\end{equation}
$\underline{Z}^{-1}$ is just the $2\times 2$ upper restriction of the
dynamical conductance matrix $\underline{G}$ found for the gated LL.
Equation (\ref{main-result3}) is the main result of this sub-section. It can
be rewritten as:

\[
\left( 
\begin{array}{c}
i_{1} \\ 
i_{2}
\end{array}
\right) =\left[ \left( 
\begin{array}{cc}
1 & 0 \\ 
0 & -1
\end{array}
\right) -\underline{{\Huge Z}}^{-1}\left( 
\begin{array}{cc}
R_{S}+i\frac{1}{\omega {\cal C}}\partial _{x} & 0 \\ 
0 & R_{D}++i\frac{1}{\omega {\cal C}}\partial _{x}
\end{array}
\right) \right] \left(
\begin{array}{c}
ej_{0}(-a) \\ 
-ej_{0}(a)
\end{array}
\right) 
\]
where ${\cal C=}\frac{2Ke^{2}}{h\,u}$ is the intrinsic LL capacitance and
where for instance 
\[
j_{0}(x)=\frac{i\omega eK}{u\pi }\int_{0}^{x}dy\int_{0}^{y}dzE(z)e^{i\frac{%
\omega }{u}(x+z-2y)}. 
\]
It is easily checked that the current response is independent of the
particular solution $j_{0}$ chosen: shifting $j_{0}$ and $\rho _{0}=\frac{-1%
}{i\omega }\partial _{x}j_{0}$ by either chiral currents $\delta j_{+}$ or $%
\delta j_{-}$ gives contributions which cancel each other.

Even for the subcase $R_{S}=R_{D}=R_{0}/2$ this matrix equation does not
seem to appear in the literature: for instance Ponomarenko writes the
current at a given position in the inhomogeneous LL model for an arbitrary
electric field as a Green function convolution\cite{3} but that simple
matrix relation between the response of a gated wire and the response to an
arbitrary electric field is not explicitly written. Joint measurements in
both context would be interesting to reveal such relations between the
dynamical responses.

\subsection{Uniform electric field.}

We now specialize the discussion to a uniform electric field $E=-\partial
xV_{3}$ so that:

\[
j_{0}=-\frac{i\,u\,Ke}{\omega \,\pi }E\,=-i2G_{0}\frac{u}{\omega }E\, 
\]
The current response is therefore: 
\begin{eqnarray*}
\left( 
\begin{array}{c}
i_{1} \\ 
i_{2}
\end{array}
\right) &=&ej_{0}\left[ \underline{{\Huge Z}}^{-1}\left( 
\begin{array}{c}
R_{S} \\ 
-R_{D}
\end{array}
\right) +\left( 
\begin{array}{c}
1 \\ 
-1
\end{array}
\right) \right] \\
&=&-iE\frac{G_{0}u\,K}{\omega }\left( 
\begin{array}{c}
\frac{\left( \frac{1}{2}+K\overline{R_{D}}\right) -\left( \frac{1}{2}-K%
\overline{R_{D}}\right) \exp i2\phi -2K\overline{R_{D}}\exp i\phi }{\left( 
\frac{1}{2}+K\overline{R_{S}}\right) \left( \frac{1}{2}+K\overline{R_{D}}%
\right) -\exp i2\phi \left( \frac{1}{2}-K\overline{R_{S}}\right) \left( 
\frac{1}{2}-K\overline{R_{D}}\right) } \\ 
-\frac{\left( \frac{1}{2}+K\overline{R_{S}}\right) -\left( \frac{1}{2}-K%
\overline{R_{S}}\right) \exp i2\phi -2K\overline{R_{S}}\exp i\phi }{\left( 
\frac{1}{2}+K\overline{R_{S}}\right) \left( \frac{1}{2}+K\overline{R_{D}}%
\right) -\exp i2\phi \left( \frac{1}{2}-K\overline{R_{S}}\right) \left( 
\frac{1}{2}-K\overline{R_{D}}\right) }
\end{array}
\right) .
\end{eqnarray*}
If $R_{S}=R_{D}=R_{0}/2$, this yields: 
\[
\left( 
\begin{array}{c}
i_{1} \\ 
i_{2}
\end{array}
\right) =-iE\frac{2G_{0}u\,K}{\omega }\frac{-i\sin \frac{\phi }{2}}{K\cos 
\frac{\phi }{2}-i\sin \frac{\phi }{2}}\left( 
\begin{array}{c}
1 \\ 
-1
\end{array}
\right) , 
\]

which is exactly the expressions found by Sablikov et al.\cite{6} and
Ponomarenko\cite{3}.

Sablikov and Shchamkhalova argue that due to a charging of the reservoirs
the real current measured in an AC experiment is not $i_{1}$ but that one
must add a displacement current $\frac{dQ_{S}}{dt}$ where $Q_{S}$ is the
charge appearing at the source\cite{6,8}. Appealing to a result initially
derived by Shockley, using Laplace equation they find that: 
\[
\frac{dQ_{S}}{dt}=-i_{1}+\frac{1}{L}\int_{-L/2}^{L/2}i(x)\,dx, 
\]
and therefore the current measured at the left electrode is: 
\[
i_{mes}=\frac{1}{L}\int_{-L/2}^{L/2}i(x)\,dx, 
\]
for {\it a uniform electric field} and {\it plane electrodes orthogonal to
the wire}.

That point of view is however valid only if one does not take into account
relaxation processes in the reservoir: the charging of the reservoir must be
taken into account only for frequencies $\omega \gg 1/\tau _{rel}$ where $%
\tau _{rel}$ is the relaxation time of the reservoir, i.e. the inverse of
the plasma frequency $\omega _{P}=1/\tau _{rel}\sim 10^{15}Hz$. For optical
processes this becomes relevant but not for the transport experiments one
considers here.

It is quite easy to extract the distribution of current and charge in the
sample: 
\[
i(x,\omega )=K\frac{e}{h}\left( \mu _{+}^{\prime }(x,\omega )-\mu
_{-}^{\prime }(x,\omega )\right) +ej_{0}(x).
\]

Since:

\begin{eqnarray*}
\mu _{+}^{\prime }(x,\omega ) &=&\exp i\frac{\phi }{2}\exp i\frac{\omega x}{u%
}\ \mu _{+}^{\prime }(-a,\omega ), \\
\mu _{-}^{\prime }(x,\omega ) &=&\exp i\frac{\phi }{2}\exp -i\frac{\omega x}{%
u}\ \mu _{-}^{\prime }(a,\omega ),
\end{eqnarray*}
it follows: 
\[
i(x)=K\frac{e}{h}\exp i\frac{\phi }{2}\left( 
\begin{array}{cc}
\exp i\frac{\omega x}{u}, & -\exp -i\frac{\omega x}{u}
\end{array}
\right) \cdot \overrightarrow{\mu }+ej_{0}(x).
\]
Using the relation between $\overrightarrow{\mu }$ and $j_{0}$ (eq.(\ref
{mu-j}) above where one takes $\rho _{0}=0$ because the electric field is
uniform) one easily finds:

\[
i(x)=ej_{0}\left[ 1-\frac{K\left( \cos (\frac{\omega x}{u})\left( \overline{%
R_{S}}+\overline{R_{D}}\right) \left( \cos \frac{\phi }{2}-i4K\frac{R_{q}}{%
R_{0}}\sin \frac{\phi }{2}\right) -\sin (\frac{\omega x}{u})\sin \frac{\phi 
}{2}\left( \overline{R_{S}}-\overline{R_{D}}\right) \right) }{\left( \frac{1%
}{2}+K\overline{R_{S}}\right) \left( \frac{1}{2}+K\overline{R_{D}}\right)
-\exp i2\phi \left( \frac{1}{2}-K\overline{R_{S}}\right) \left( \frac{1}{2}-K%
\overline{R_{D}}\right) }\right] 
\]

For the symmetric case $R_{S}=R_{D}=R_{0}/2$ this reduces to: 
\[
i(x)=ej_{0}\left[ 1-\frac{K\cos (\frac{\omega x}{u})}{K\cos \frac{\phi }{2}%
-i\sin \frac{\phi }{2}}\right] ,
\]
which is also found by Sablikov et al.\cite{6}. 

The density is then easily found as $\rho (x)=-\frac{1}{i\omega }\partial
_{x}i(x)$.

\section{Conclusions.}

In this paper we have discussed consequences on AC transport of the
inclusion of arbitrary interface resistances $R_{S}$ and $R_{D}$ between the
sample and the source and drain electrodes. The resistive coupling of the
Luttinger liquid to the electrodes is described using a boundary condition
formalism.

We considered a gated two-port Luttinger liquid which enabled us to
generalize expressions of the dynamical conductance matrix. By considering
the dynamical impedance we were able in particular to show that in the
low-frequency limit the Luttinger liquid can be modelled as an electrical
circuit comprising an inductance per unit length ${\cal L}=\frac{h}{2u\
Ke^{2}}$ in series with the interface resistances, the whole being
capacitively coupled to the ground with intrinsic conductance per unit
length ${\cal C=}\frac{2Ke^{2}}{h\,u}$. That transmission line analogy is
however invalid beyond order two in frequency.

Focusing in the impedance response of the LL we showed that a joint
measurement of both dynamical impedance and gate conductance $G_{33}$ up to
order one in frequency is sufficient to extract the Luttinger parameters. A
measurement up to order two allows extraction of the interface resistances
whose quantization can therefore be checked (or disproved).

We then considered the application of an arbitrary AC electric field along
the sample; we then discussed the case of a uniform electric field
generalizing earlier results valid only for $R_{S}=R_{D}=R_{0}/2$ . The
author acknowledges useful discussions with C. Texier and H\'{e}l\`{e}ne
Bouchiat's group.

\begin{figure}[tbp]
\includegraphics[scale=1]{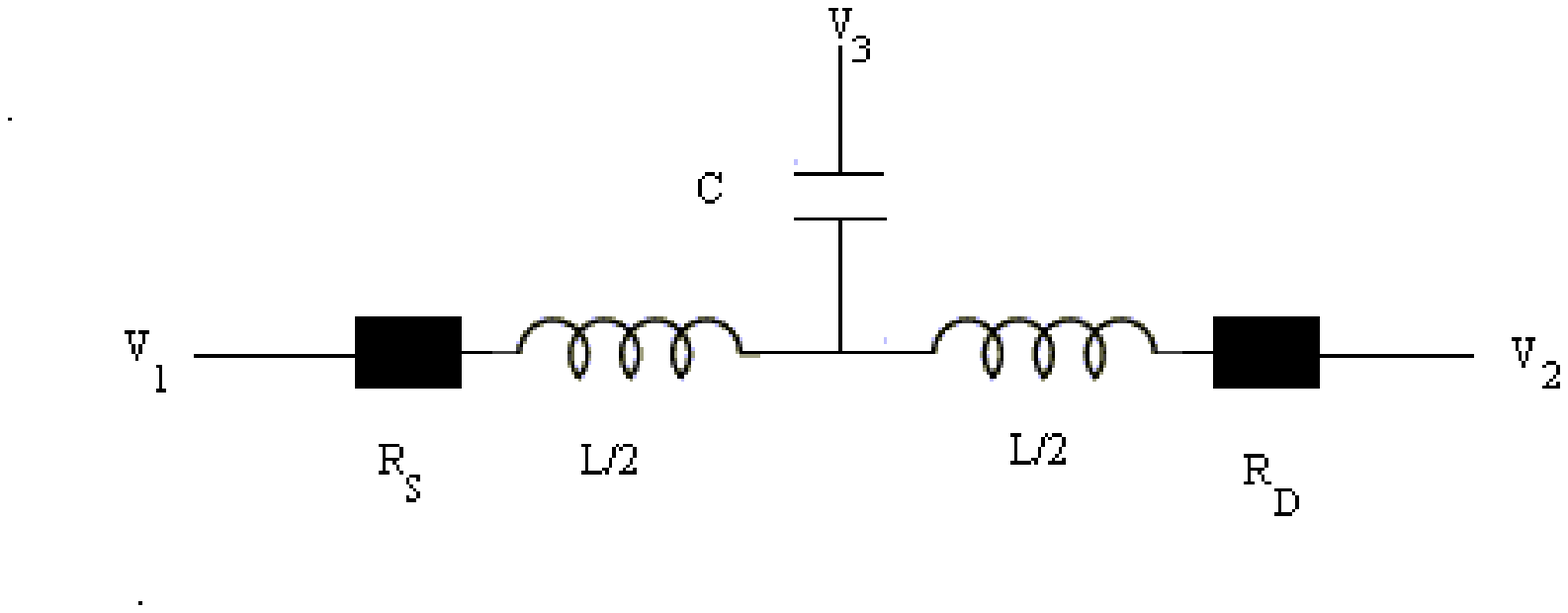}
\caption{Electrical circuit equivalent to the Luttinger liquid. The
inductance per unit length is ${\cal L=}\frac{h}{2u\,Ke^{2}}$ and the
capacitance per unit length is ${\cal C}=\frac{2Ke^{2}}{h\,u}$.}
\end{figure}


\begin{thebibliography}{99}
\bibitem{1}  F. D. M. Haldane, J. Phys. C 14, 2585 (1981); H.J. Schulz, in
Les Houches Summer School 1994, Mesoscopic Quantum Physics, E. Akkermans et
al. eds, Elsevier Science, Amsterdam (1995); M. P. A. Fisher, L. I. Glazman,
in Mesoscopic Electron Transport, L. Kouwenhoven et al eds, Kl\"{u}wer
Academic Press, Dordrecht (1997); on fractionalization: K.-V. Pham, M.
Gabay, P. Lederer, Phys. Rev. B 61, 16397 (2000).

\bibitem{2}  S. Tarucha, T. Honda, T. Saku, Solid State Commun. 94, 413
(1995); R. Egger, A. Bachtold, M. Fuhrer, M. Bockrath, D. Cobden, P. McEuen,
in Interacting electrons in nanostructures, P. Haug, H. Schoeller eds,
Springer, Berlin (2001).

\bibitem{3}  V. V. Ponomarenko, Phys. Rev. B 54, 10328 (1996).

\bibitem{4}  Ya. M. Blanter, F. W. Hekking, M. B\"{u}ttiker, Phys. Rev.
Lett. 81, 1925 (1998).

\bibitem{5}  I. Safi, Ann. Phys. (Paris) 22, 463 (1997); Eur. Phys. J. B 12,
451 (1999).

\bibitem{6}  V. A. Sablikov, B. S. Shchamkhalova, JETP Lett. 66, 41 (1997).

\bibitem{7}  G. Cuniberti, M. Sassetti, B. Kramer, J. Phys. Condens. Matter
8, L21 (1996); Phys. Rev. B 57, 1515 (1998).

\bibitem{8}  V. A. Sablikov, B. S. Shchamkhalova, Phys. Rev. B 58, 13847
(1998).

\bibitem{9}  P. J. Burke, cond-mat/0204262 and cond-mat/0207222, submitted
to IEEE.

\bibitem{10}  M. W. Bockrath, Ph. D. Thesis, University of California,
Berkeley, (1999).

\bibitem{11}  E. B. Sonin, J. Low Temp. Phys. 1, 321 (2001); R. Tarkiainen
et al, Phys. Rev. B. 64, 195412-1 (2001).

\bibitem{12}  Y. Imry, in Directions in Condensed Matter Physics, ed. by G.
Grinstein and G. Mazenko, World Scientific, Singapore, 1986.

\bibitem{13}  M. B\"{u}ttiker, Phys. Rev. B 40, 3409 (1989).

\bibitem{14}  K-V Pham, F. Pi\'{e}chon, K-I Imura, P. Lederer,
cond-mat/0207294.

\bibitem{15}  L. I. Glazman, I. M. Ruzin, B. I. Shklovskii, Phys. Rev. B 45,
8454 (1992).

\bibitem{16}  M. B\"{u}ttiker, H. Thomas, A. Pr\^{e}tre, Phys. Lett. A 180,
364 (1993); M. B\"{u}ttiker, J. Korean Phys. Soc. 34, 121-130 (1999).
\end{thebibliography}
\end{document}